# Impact of Spatial Dimension on Structural Ordering in Metallic Glass

Yuan-Chao Hu,[1,2,*] Hajime Tanaka[3] and Wei-Hua Wang[1,2,†]

[1]*Institute of Physics, Chinese Academy of Sciences, Beijing 100190, China*

[2]*University of Chinese Academy of Sciences, Beijing 100049, China*

[3]*Department of Fundamental Engineering, Institute of Industrial Science, University of Tokyo, 4-6-1 Komaba, Meguro-ku, Tokyo 153-8505, Japan*

*ychu0213@gmail.com; †whw@iphy.ac.cn

Metallic glasses have so far attracted considerable attention for their applications as bulk materials. However, new physics and applications often emerge by dimensional reduction from three dimension (3D) to two dimension (2D). Here, we study, by molecular dynamics simulations, how the liquid-to-glass transition of a binary $Cu_{50}Zr_{50}$ MG is affected by spatial dimensionality. We find clear evidence that crystal-like structural ordering controls both dynamic heterogeneity and slow dynamics, and thus plays a crucial role in the formation of the 2DMG. Although the 2DMG reproduces the dynamical behaviors of its 3D counterpart by considering Mermin-Wagner-type fluctuations specific to 2D, this atomic-scale structural mechanism is essentially different from that for the 3DMG in which icosahedral clusters incompatible with crystallographic symmetry play a key role in glassy behaviors. Our finding provides a new structural mechanism for the formation of 2DMGs, which cannot be inferred from the knowledge of 3DMGs. The results suggest a structural basis for the glass transition in 2DMG and provide possible explanations for some previous experimental observations in ultrathin film MGs.

## I. INTRODUCTION

Striking breakthroughs toward new physics and applications are often made by the reduction of spatial dimensionality of materials from three dimensions (3D) to (quasi-) two-dimensions (2D). Some distinguished examples are quantum Hall effect, topological and structural phase transitions, superconductivity, melting mechanism and ultra-stable glasses [1-4]. Recently, the general impact of spatial dimensionality on the glass transition phenomena, one of the deepest and most controversial unsolved problem in condensed matter physics and materials sciences, has also been actively debated from a viewpoint of fluctuations specific to low-dimensional systems [5-9]. For atomic metallic glass-formers, main attentions have so far been paid to the bulk systems due to their promising applications as structural materials. Thus, computer simulations are primarily devoted to establishing the structure-property relationship in 3D metallic glasses (3DMGs) [10]. There have been a large number of simulation studies on 2D glass-formers, but which have been limited to model systems only: there have so far been few researches addressing a question of how MGs can form in 2D. It is quite interesting to ask what atomic-scale structural mechanism leads to the formation of 2DMGs and what is the similarity and difference between 2D and 3DMGs. Answers to these fundamental questions remain unknown up to now. Answering them not only could assist in understanding some long-standing fundamental issues such as glass transition and formation and glass-forming ability (GFA) of MGs, but also is of practical significance to promote functional applications of quasi-2D thin film MGs in microelectromechanical systems, biomedical and energy-related areas [11,12].

Addressing the above problems in experiments is currently difficult because of the limited microscopic resolution to track atom motions and the difficulties in synthesizing 2DMGs. In this work, we try to fabricate a realistic model 2DMG and investigate the liquid-to-glass transition in detail at an atomistic scale by molecular dynamics simulations to answer the above questions. Crystallization happens inevitably for all metallic glass-forming liquids if they were cooled slower than their critical cooling rates, i.e. meeting the *nose* in the time-temperature-transformation (TTT) diagram [13]. It is known that some alloys are easy to vitrify and have good GFA while others not. The system we considered here is $Cu_{50}Zr_{50}$, which is known to almost have the best GFA in 3D among binary metallic systems. Unlike its 3D system where icosahedral clusters incompatible with crystallographic symmetry are believed to play a key role in the glassy behaviors, here we found that crystal-like structural ordering controls both dynamic heterogeneity and slow dynamics in the 2D system, and thus plays a crucial role in the formation of the 2DMG. This finding demonstrates that there can be a significant influence of spatial dimensionality on the structural mechanism of glass transition in MG.

## II. METHODS

We performed molecular dynamics simulations using the Large-scale Atomic/Molecular Massively Parallel Simulator (LAMMPS) [14]. The optimized embedded-atom method (EAM) potential is used to describe the interatomic interactions [15]. This semi-empirical potential is developed by combining experimental characterization data and results from first-principle calculations. It aims to provide suitable description for the liquid and amorphous Cu-Zr alloys. More details about the potential are available in Ref. [15]. To avoid large fluctuations in pressure in 2D, we carried out the standard canonical ensemble (NVT) simulations over a range of temperatures with periodic boundary conditions. Here we note that although our simulations were carried out in the NVT ensemble the results will not be altered by an ensemble used. The number density $\rho$ designed is 0.15 $\text{Å}^{-2}$ and the atom number $N$ is 10 000. The time step $dt$ used for integration is 0.001 and 0.002 ps. The temperature is maintained by the *Nośe-Hoover* thermostat. To characterize the GFA of Cu$_{50}$Zr$_{50}$ in 2D, we quenched the equilibrium metallic liquid at 1500 K to 1.0 K continuously at different cooling rates and checked whether first order phase transition happens or not in these processes. For all the simulations in the supercooled liquid states, the sample was first equilibrated for at least $100\tau_\alpha$ at the target temperature and then the production runs were carried out also for at least $100\tau_\alpha$. 10-15 independent simulations were performed for each state point for ensemble average.

## III. RESULTS AND DISCUSSION

We successfully fabricated this binary MG in 2D and performed molecular dynamics simulations in canonical ensemble. **Figure 1(a)** shows the temperature (*T*) dependence of the atomic potential energy $\langle U \rangle$ at several cooling rates. The lowest cooling rate is $10^9$ K s$^{-1}$ and the cooling time is 1499 ns. This is almost the lowest cooling rate ever accessed in simulating MGs by continuous cooling. The absence of a discontinuous jump in the $\langle U \rangle - T$ curves unambiguously demonstrates the good GFA of the 2DMG, at least in the simulation accessible timescale. The glass transition shifts to lower temperatures for smaller cooling rates [16].

To unveil how the 2DMG forms, we studied the liquid-to-glass transition and characterized its structural and dynamical properties during cooling. The overall static structure of the supercooled 2DMG is characterized by the radial distribution functions $g(r) = \frac{1}{2\pi r \Delta r \rho (N-1)} \langle \sum_{j \neq k} \delta(r - |\vec{r}_{jk}|) \rangle$ and the static structure factors $S(q) = \frac{1}{N} \langle \sum_k \sum_j e^{-i\vec{q} \cdot (\vec{r}_k - \vec{r}_j)} \rangle$ where $\vec{r}_k$ is the positional vector of atom *k* and $i = \sqrt{-1}$. The changes in $g(r)$ and $S(q)$ upon reducing temperatures are shown in **Fig.**

**1(b)** and **1(c)**, respectively. Both $g(r)$ and $S(q)$ include information on two-point density-density correlation. The quick decay of $g(r)$ indicates the absence of long-range translational order in the supercooled 2DMG while the splitting of the second peak at lower temperatures suggests the emergence of some structural ordering at medium-range [17]. Thus, the supercooled 2DMG shows typical static structural features of metallic glass-forming liquids. Compared to the complex local coordination structures in 3DMGs [18,19], the fluctuations of the coordination number $Z$ of an atom in 2D is small: $Z$~5-7. The neighbors of each atom are defined by the cutoff distances equal to the positions of the first minimums in the partial $g(r)$. Taking advantage of easy visualization in 2D, we show the distribution of $Z$ in **Fig. 2(a)**, in which atoms mainly have 6 nearest neighbors resulting from denser packing and lower configurational (or, higher vibrational) entropy. There are also dispersed topological defects with $Z \neq 6$, which are crucial to frustrate crystallization [20]. This frustration may be regarded to be of entropic origin. To probe the characteristic feature of the local structure, we employed the hexatic order parameter $\psi_6^j = \frac{1}{n_j}\sum_{m=1}^{n_j} e^{i6\theta_m^j}$ where $\theta_m^j$ is the angle between $(\vec{r}_m - \vec{r}_j)$ and the $x$-axis, and $n_j$ is the number of the nearest neighbors of atom $j$. We then defined the parameter $\Psi_6^j$ by averaging $\psi_6^j$ over the relative structural relaxation time $\tau_{\alpha,R}$ measured through the overlap function (see below), $\Psi_6^j = \frac{1}{\tau_{\alpha,R}}\int_{t_0}^{t_0+\tau_{\alpha,R}} dt|\psi_6^j|$, to characterize a local structure. $\Psi_6^j = 1$ means the perfect hexagonal arrangement of six nearest neighbor atoms around particle $j$ while $\Psi_6^j = 0$ represents a random rearrangement. Time-averaging is only to remove short-time fluctuations in the real space representation and is not physically essential for equilibrium supercooled liquids. **Figure 2(b)** is an exemplified snapshot of the spatial distribution of $\Psi_6^j$ at 670 K. It is evident that some spatially extending crystal-like structural ordering develops heterogeneously in 2DMG, which are anti-correlated with the topological defects shown in **Fig.**

**2(a)**. The spatial correlation of $\psi_6$ is calculated as $g_6(r) = \frac{1}{2\pi r \Delta r \rho (N-1)} \langle \sum_{j \neq k} \delta(r - |\vec{r}_{jk}|) \psi_6^j \psi_6^{k*} \rangle$ where * represents the complex conjugate. The structural correlation length $\xi_6$ is then evaluated by fitting the envelope of $g_6(r)/g(r)$ to the 2D Ornstein-Zernike (OZ) function $\sim r^{-1/4} \exp(-r/\xi_6)$, as illustrated in **Fig. 2(c)** [21]. The division by $g(r)$ is to remove the effect of short-range translational ordering. The exponential-like decay of $g_6(r)/g(r)$ indicates the absence of long-range orientational order in the 2DMG and further eliminates a possibility of crystallization. Here we emphasize that the local crystal-like bond orientational ordering should not be confused with crystal nuclei. The absence of crystals can also be confirmed by the fact that the ordering is not associated with any density change, which is evidenced by the absence of excess scattering in the low $q$ region of S($q$) [**Fig. 1(c)**]. In addition, when averaged over a longer time, the structural heterogeneity associated with the hexatic order disappears (not shown here). This indicates that this heterogeneity has a finite lifetime and does not grow with time. The lifetime and spatial extent of the local crystal-like orientational order both increase with decreasing temperature [**Fig. 2(c)**]. When comparing the cage relative mobility field in **Fig. 2(d)** with the static structure in **Fig. 2(b)**, we can clearly see that atoms with higher degree of local crystal-like ordering move slower, or particles in more disordered regions are more mobile. This strongly demonstrates a link of static heterogeneity in local crystal-like orientational order to dynamic heterogeneity.

In order to further confirm the connection between the structure and dynamics, we adopted a statistical method - isoconfigurational ensemble [22]. We performed 100 simulations starting from the same initial configuration but with different velocity distributions. The local structure was then characterized by using $\Psi_6^j$ from all simulations over a time $\tau_{\alpha,R}/10$, so the structural information $\langle \Psi_6^i \rangle_{\text{iso}}$ was obtained through isoconfigurational average, as visualized in **Fig. 3(a)** [23]. We also evaluated the corresponding dynamical properties in isoconfigurational ensemble by local Debye-Waller factor and dynamic propensity, which are associated with the probability of an atom in the initial configuration undergoing a relative displacement within time intervals of $\tau_{\alpha,R}/100$ and $\tau_{\alpha,R}$, respectively. The local Debye-Waller factor characterizes the short-time fluctuations when atoms are rattling in the cages formed by their neighbors (fast $\beta$ relaxation) whereas dynamic propensity is related to the cage-break motion in the diffusive regime ($\alpha$ relaxation). When comparing the isoconfigurational averaged structure and dynamics in **Fig. 3(a)-3(c)**, atoms with higher crystal-like ordering have larger solidity and are less mobile. The correspondence between structure and dynamics is impactful not only for the fast $\beta$ relaxation, but also for the $\alpha$ relaxation. Besides the established relationship between short-time and long-time dynamic heterogeneities [22], we make further efforts in finding out the structural precursor of the subsequent inhomogeneous dynamics. The direct atomic-scale correspondences have never been reported for MGs in 3D. Particularly interesting is the fact that for 3D bulk $Cu_{50}Zr_{50}$ only icosahedral clusters incompatible with crystallographic symmetry has been considered to be responsible for slow glassy dynamics. This indicates that a different selection mechanism of structural ordering works between 2D and 3D at least for CuZr, which has a crucial impact on the material design of low dimensional MGs.

Next we show the time-space correlation in the 2DMG. The most prominent feature of glass transition is the dramatic increase of the structural relaxation time $\tau_\alpha$ when $T$ decreases moderately. Unveiling the glass formation mechanism is almost equivalent to understanding how and why $\tau_\alpha$ grows drastically. To evaluate the relaxation times, we considered the decay of an overlap function $Q(t) = \sum_{j=1}^{N} \omega(|\vec{r}_j(0) - \vec{r}_j(t)|)$ where $\omega(r) = 1$ if $r \leq a$ and zero otherwise. The threshold $a$ was set to be 1.0 Å. $Q(t)$ characterizes the number of atoms moving less than a distance over time $t$ [24]. $\tau_\alpha$ is determined when $\langle Q(t) \rangle / N$ decays to $e^{-1}$. In addition to the standard dynamic quantities evaluated from the absolute displacement $|\Delta\vec{r}_i|$ of atom $i$, we also measured the dynamics using the displacement $|\Delta\vec{R}_i|$ relative to the cage to isolate the diffusive motion from low-frequency vibrational modes, which are Mermin-Wagner-type fluctuations specific to 2D systems [7-9,23]. The relative displacement $|\Delta\vec{R}_i|$ is defined as $|\vec{R}_i(t) - \vec{R}_i(0)|$ of atom $i$ with respect to its $n_i$ nearest neighbors: $\vec{R}_i = \vec{r}_i - (1/n_i)\sum_j^{n_i}\vec{r}_j$. Therefore, the cage relative overlap function $\langle Q_R(t)\rangle$ (a subscript R means a relative quantity) is defined as $Q_R(t) = \sum_{j=1}^{N}\omega(|\vec{R}_j(0) - \vec{R}_j(t)|)$ and the relative structural relaxation times $\tau_{\alpha,R}$ are also determined when $\langle Q_R(t)\rangle/N$ decays to $e^{-1}$. This strategy is also known to be effective to reduce the strong finite-size effects in 2D [6,7,9]. We show the results of the dynamics obtained from $|\Delta\vec{R}_i|$ in the manuscript. By fitting $\tau_{\alpha,R}$ to the Vogel-Fulcher-Tammann relation

$$\tau_{\alpha,R} = \tau_{0,R}\exp[D_R T_{0,R}/(T - T_{0,R})], \qquad (1)$$

the fragility index $D_R$ and the ideal glass transition point $T_{0,R}$ are determined to be around 4.65 and 420 K, respectively. Thus, we may say that the binary 2DMG is a typical fragile glass-former [25]. The spatial heterogeneity of the dynamics was then characterized by a relative four-point dynamic susceptibility $\chi_{4,R}(t) = N^{-1}[\langle Q_R(t)^2\rangle - \langle Q_R(t)\rangle^2]$ and a relative four-point structure factor of *immobile* atoms $S_{4,R}(q;t) = \frac{1}{N}[\langle W_R(\vec{q},t)W_R(-\vec{q},t)\rangle - \langle W_R(\vec{q},t)\rangle\langle W_R(-\vec{q},t)\rangle]$ where $W_R(\vec{q};t) = \sum_{j=1}^{N}\exp[i\vec{q}\cdot\vec{r}_j(0)]\omega(|\vec{R}_j(t) - \vec{R}_j(0)|)$ [24,26]. The $T$-dependence of $\chi_{4,R}(t)$ in **Fig. 4(a)** shows a peak at an intermediate timescale $\tau_p$ that increases with reducing temperature, manifesting growing dynamic heterogeneity. To evaluate the dynamic correlation length $\xi_{4,R}$, we calculated $S_{4,R}(q;t)$ at $\tau_p$ [24]. As shown in **Fig. 4(b)**, $\xi_{4,R}$ is determined by fitting $S_{4,R}(q;\tau_p)$ to the OZ function $S_{4,R}(q;\tau_p) = S_{4,R}(q=0;\tau_p)/\left[1 + (q\xi_{4,R})^2\right]$ in the low $q$ region. In the scaled plot of $S_{4,R}(q;\tau_p)/S_{4,R}(q=0;\tau_p)$ versus $q\xi_{4,R}$ (see the inset), all data can be collapsed. Here we should mention that the four-point dynamic correlation length would be overestimated if we do not consider the influence of Mermin-Wagner-type fluctuations in 2D. It is striking that $\xi_{4,R}$ is proportional to $\xi_6$ in the supercooled 2DMG and $\xi_{4,R}/\xi_{4,R0} \cong \xi_6/\xi_{60}$, as depicted in **Fig. 4(c)**. The hand-by-hand growth of the structural and dynamical correlation lengths on cooling shares the same trend through the power-law relation

$$\xi = \xi_0[(T-T_{0,R})/T_{0,R}]^{-\nu} \ (\xi = \xi_6; \xi_{4,R}), \qquad (2)$$

where $T_{0,R}$ is determined independently by Eq. (1), $\xi_0$ and $\nu$ are the free fitting parameters. We obtain $\nu \cong 1.0$. This behavior strongly implies that dynamic heterogeneity in 2DMG originates from the fluctuations of the crystal-like structural ordering, which is quite different from the 3D counterpart [10,19]. The value of $\nu$ ($\cong 1.0$) is equal to the exponent of the correlation length in the Ising universality class, $d/2$, where $d$ is the spatial dimension, indicating that critical-like fluctuations belonging to the Ising universality class may be behind the glass transition [21,27]. To find out what controls the drastic growth of $\tau_{\alpha,R}$, we combine Eq. (1) and Eq. (2) to reach the following empirical relation

$$\tau_{\alpha,R} = \tau_{0,R}\exp[D_R(\xi/\xi_0)] \ (\xi = \xi_6; \xi_{4,R}). \qquad (3)$$

We show the comparison of the data with Eq. (3) in **Fig. 4(d)**. We note that this comparison requires no fitting parameters. This relation is consistent with the predictions of the two-order parameter (TOP) model [28] and random first-order transition (RFOT) theory [29]. Here we note that the coherent growth of static and dynamic correlation length is more consistent with the former than the latter, since in the RFOT theory the metastability is necessary to have static correlation [29] and thus the dynamic correlation is predicted to grow faster than the static one upon cooling.

The $\log\tau \sim \xi$ scaling relation between dynamics and structure further indicates that the crystal-like structural ordering may be the origin of dynamic heterogeneity and slow dynamics in this 2DMG. To our knowledge, such a relation has never been reported for MGs.

We propose a possible atomic-scale structural mechanism of how the MG forms in 2D, based on the above results of the formation of 2DMG and the liquid-to-glass transition behavior. While crystallization is frustrated by dispersed topological defects with five- or seven-fold symmetry, which may originate from entropic effects, the growth of the correlation length of transient local crystal-like order with lower mobility upon cooling directly leads to that of the dynamic correlation length, i.e. dynamic heterogeneity. Following the growing length and lifetime of the static correlation, the energy barrier of local cooperative rearrangements increases rapidly, resulting in the dramatic slowdown. The growth of structural order also accompanies the decrease of the configurational entropy. This atomic structural mechanism of 2DMG formation is distinct from the current understanding of its 3D counterpart and some other 3DMGs, in which the formation of icosahedral clusters is believed to govern the glass transition [10,19,30]. Our finding also indicates that although the glassy dynamics of the 2DMG is basically the same as that of its 3D counterpart besides the presence of Mermin-Wagner-type fluctuations specific to 2D, as shown in Refs. [7-9], the nature of structural order (spatial extendability) behind slow glassy dynamics can be quite different in 2D and 3D, at least for the CuZr system studied here.

The findings are of both theoretical and practical significance to understand MGs. On one hand, the results suggest a structural basis for the glass transition in 2D and imply a causal link between vitrification and crystallization [28], compatible with the previous findings in colloidal and granular liquids [20,21]. In the scenario of TOP, vitrification is the result of frustration on the way to crystallization [20,21,28]. This is also reasonable in 2DMGs since the thermodynamic ground state is crystal but frustration effects strongly impede crystallization during cooling. On the other hand, the formation mechanism of 2DMGs is vital to understand the properties of quasi-2D thin film MGs. Although this crystal-like ordering is hidden in scattering experiments since the order cannot be detected by two-body density correlation, it is measurable in sophisticated experiments by high-resolution transmission electron microscopy (HRTEM) and fluctuation electron microscopy [11,31-34]. This frozen order in disorder is probably the reason why tiny regions of crystal-like order have been detected by HRTEM in ultra-thin amorphous film [34] and in as-deposited quasi-2D Zr-, Fe- and Cu-based thin film MGs [11]. These results, in turn, validate that these orderings are intrinsic because no crystallization shall happen in these good glass-formers at high cooling rates in deposition. Due to the low interfacial energy between a region of a liquid with crystal-like ordering and a crystal, the crystal-like order formed in a supercooled liquid could act as a precursor for crystallization [34,35]. This indicates that the ordering should play a crucial role in the thermal stability of (quasi-) 2DMGs. Such structural ordering would also influence the mechanical properties of the amorphous solids.

## IV. SUMMARY

In summary, by simulating a binary MG in 2D, we found that local crystal-like orientational ordering is closely linked to both dynamic heterogeneity and slow dynamics, and thus facilitate formation of the 2DMG. This is markedly different from the currently believed structural mechanism in its 3D counterpart. This difference in the type of structural order may not be general among all MGs since in some 3DMGs tiny crystal-like clusters have been detected in experiments [31-33]. Crystal-like structural orderings likely take place in some 3DMGs, on noting that our results are consistent with previous findings in experiments and simulations of metallic and non-metallic glass-formers [21,31-33,36,37]. However, our study clearly indicates that for some systems such as CuZr the spatial dimension can seriously affect the type of structural ordering favored in a supercooled state and thus the glass-formation mechanism. Quite interestingly, our study shows that the dimensional reduction from 3D to 2D in the studied system $Cu_{50}Zr_{50}$ leads to the change in locally favored particle configuration from *non-crystallographic* isosahedral-like structures to *crystal-like* hexagonal structures. This finding suggests that we cannot apply a knowledge in 3D to 2D systems in a straightforward manner. It would be quite interesting to study how the nature of structural ordering in MGs changes in the case of thin film as a function of film thickness using the protocol of Ref. [38]. For example, it is expected that the GFA and fragility would change as a function of the film thickness. Thus, our work is also of practical significance in understanding and designing the properties of quasi-2D thin film MGs.

## ACKNOWLEDGEMENTS

We thank Prof. Y. Yang and Prof. P. F. Guan for helpful discussions. The work was supported by the NSF of China (51571209 and 51461165101), the MOST 973 Program (2015CB856800) and the Key Research Program of Frontier Sciences, CAS (QYZDY-SSW-JSC017). H.T. acknowledges a support by Grants-in-Aid from Specially Promoted Research (KAKENHI Grant No. 25000002) from the Japan Society for the Promotion of Science (JSPS).

# References

[1] J. M. Kosterlitz and D. J. Thouless, J. Phys. C **6**, 1181 (1973).

[2] C. Z. Chang, J. Zhang, X. Feng, J. Shen, Z. Zhang, M. Guo, K. Li, Y. Ou, P. Wei, L. L. Wang, Z. Q. Ji, Y. Feng, S. Ji, X. Chen, J. Jia, X. Dai, Z. Fang, S. C. Zhang, K. He, Y. Wang, L. Lu, X. C. Ma, and Q. K. Xue, Science **340**, 167 (2013).

[3] S. He, J. He, W. Zhang, L. Zhao, D. Liu, X. Liu, D. Mou, Y. B. Ou, Q. Y. Wang, Z. Li, L. Wang, Y. Peng, Y. Liu, C. Chen, L. Yu, G. Liu, X. Dong, J. Zhang, C. Chen, Z. Xu, X. Chen, X. Ma, Q. Xue, and X. J. Zhou, Nat. Mater. **12**, 605 (2013).

[4] S. F. Swallen, K. L. Kearns, M. K. Mapes, Y. S. Kim, R. J. McMahon, M. D. Ediger, T. Wu, L. Yu, and S. Satija, Science **315**, 353 (2007).

[5] P. W. Anderson, Science **267**, 1615 (1995).

[6] E. Flenner and G. Szamel, Nat. Commun. **6**, 7392 (2015).

[7] H. Shiba, Y. Yamada, T. Kawasaki, and K. Kim, Phys. Rev. Lett. **117**, 245701 (2016).

[8] S. Vivek, C. P. Kelleher, P. M. Chaikin, and E. R. Weeks, Proc. Natl. Acad. Sci. U.S.A. **114**, 1850 (2017).

[9] B. Illing, S. Fritschi, H. Kaiser, C. L. Klix, G. Maret, and P. Keim, Proc. Natl. Acad. Sci. U.S.A. **114**, 1856 (2017).

[10] Y. Q. Cheng and E. Ma, Prog. Mater. Sci. **56**, 379 (2011).

[11] J. P. Chu, J. S. C. Jang, J. C. Huang, H. S. Chou, Y. Yang, J. C. Ye, Y. C. Wang, J. W. Lee, F. X. Liu, P. K. Liaw, Y. C. Chen, C. M. Lee, C. L. Li, and C. Rullyani, Thin Solid Films **520**, 5097 (2012).

[12] Y. C. Hu, Y. Z. Wang, R. Su, C. R. Cao, F. Li, C. W. Sun, Y. Yang, P. F. Guan, D. W. Ding, Z. L. Wang, and W. H. Wang, Adv. Mater. **28**, 10293 (2016).

[13] W. L. Johnson, Prog. Mater. Sci. **30**, 81 (1986).

[14] S. Plimpton, J. Comput. Phys. **117**, 1 (1995).

[15] M. I. Mendelev, M. J. Kramer, R. T. Ott, D. J. Sordelet, D. Yagodin, and P. Popel, Philos. Mag. **89**, 967 (2009).

[16] P. G. Debenedetti and F. H. Stillinger, Nature **410**, 259 (2001).

[17] D. Ma, A. D. Stoica, and X. L. Wang, Nat. Mater. **8**, 30 (2009).

[18] H. Sheng, W. Luo, F. Alamgir, J. Bai, and E. Ma, Nature **439**, 419 (2006).

[19] A. Hirata, L. Kang, T. Fujita, B. Klumov, K. Matsue, M. Kotani, A. Yavari, and M. Chen, Science **341**, 376 (2013).

[20] H. Shintani and H. Tanaka, Nat. Phys. **2**, 200 (2006).

[21] H. Tanaka, T. Kawasaki, H. Shintani, and K. Watanabe, Nat. Mater. **9**, 324 (2010).

[22] A. Widmer-Cooper and P. Harrowell, Phys. Rev. Lett. **96**, 185701 (2006).

[23] J. Russo and H. Tanaka, Proc. Natl. Acad. Sci. U.S.A. **112**, 6920 (2015).

[24] N. Lačević, F. W. Starr, T. B. Schrøder, and S. C. Glotzer, J. Chem. Phys. **119**, 7372 (2003).

[25] C. A. Angell, Science **267**, 1924 (1995).

[26] Y. C. Hu, P. F. Guan, Q. Wang, Y. Yang, H. Y. Bai, and W. H. Wang, J. Chem. Phys. **146**, 024507 (2017).

[27] J. S. Langer, Rep. Prog. Phys. **77**, 042501 (2014).

[28] H. Tanaka, Eur. Phys. J. E **35**, 1, 113 (2012).

[29] T. R. Kirkpatrick, D. Thirumalai, and P. G. Wolynes, Phys. Rev. A **40**, 1045 (1989).

[30] Y. C. Hu, F. X. Li, M. Z. Li, H. Y. Bai, and W. H. Wang, Nat. Commun. **6**, 8310 (2015).

[31] Y. Hirotsu, T. G. Nieh, A. Hirata, T. Ohkubo, and N. Tanaka, Phys. Rev. B **73**, 012205 (2006).

[32] Q. Wang, C. T. Liu, Y. Yang, J. B. Liu, Y. D. Dong, and J. Lu, Sci. Rep. **4**, 4648 (2014).

[33] J. Hwang, Z. Melgarejo, Y. Kalay, I. Kalay, M. Kramer, D. Stone, and P. Voyles, Phys. Rev. Lett. **108**, 195505 (2012).

[34] C. R. Cao, K. Q. Huang, N. J. Zhao, Y. T. Sun, H. Y. Bai, L. Gu, D. N. Zheng, and W. H. Wang, Appl. Phys. Lett. **105** (2014).

[35] T. Kawasaki and H. Tanaka, Proc. Natl. Acad. Sci. U.S.A. **107**, 14036 (2010).

[36] U. R. Pedersen, T. B. Schrøder, J. C. Dyre, and P. Harrowell, Phys. Rev. Lett. **104**, 105701 (2010).

[37] J. Zemp, M. Celino, B. Schönfeld, and J. F. Löffler, Phys. Rev. Lett. **115**, 165501 (2015).

[38] B. Böddeker and H. Teichler, Phys. Rev. E **59**, 1948 (1999).

# Figure Captions

**Figure 1**. Formation of 2DMG. (a) $T$-dependence of the atomic potential energy during cooling at different cooling rates. The absence of a distinct jump demonstrates glass formation in 2DMG. (b) and (c) are the radial distribution functions g($r$) and the structure factors S($q$) during cooling. The temperature range in (b) and (c) is 670, 700, 800, 900 and 1000K. The data have been shifted for clarity.

**Figure 2**. Atomic-scale correlation between structure and dynamics. (a) spatial distribution of the coordination number Z at $t_0$ at 670 K; (b) spatial distribution of $\Psi_6^j$ of the same system in (a); (c) decaying behavior of $g_6(r)/g(r)$ at different temperatures, with 2D OZ fits to the envelope (solid lines); (d) spatial distribution of the atomic relative mean-square displacements $\langle \Delta R^2(\tau_{\alpha,R}) \rangle$ for the same system in (a) and (b).

**Figure 3**. Structure-dynamics correlation in isoconfigurational ensemble. (a-c) are the isoconfigurational averaged local crystal-like orientational ordering $\langle \Psi_6^i \rangle_{iso}$, local Debye-Waller factor $\langle \Delta R^2(\tau_{\alpha,R}/100) \rangle_{iso}$ and dynamic propensity $\langle \Delta R^2(\tau_{\alpha,R}) \rangle_{iso}$ for the same system at 670 K, respectively. The displacement is measured by using relative motion over $\tau_{\alpha,R}$.

**Figure 4**. Time-space correlation in 2DMG with cage-relative displacements. (a) $T$-dependence of the relative dynamic susceptibilities $\chi_{4,R}(t)$; (b) $T$-dependence of the relative four-point dynamic structure factors fitted to the OZ function (solid lines). The inset shows the data collapse for the OZ fits; (c) consistent growth of the dynamic ($\xi_{4,R}$) and static ($\xi_6$) correlation lengths on cooling. It is obvious that $\xi_{4,R}/\xi_{4,R0} \cong \xi_6/\xi_{60}$ where $\xi_{4,R0} \cong 10.15$ and $\xi_{60} \cong 5.42$. The data are fitted by Eq. (2); (d) scaling relationship between timescale and length scales through Eq. (3).

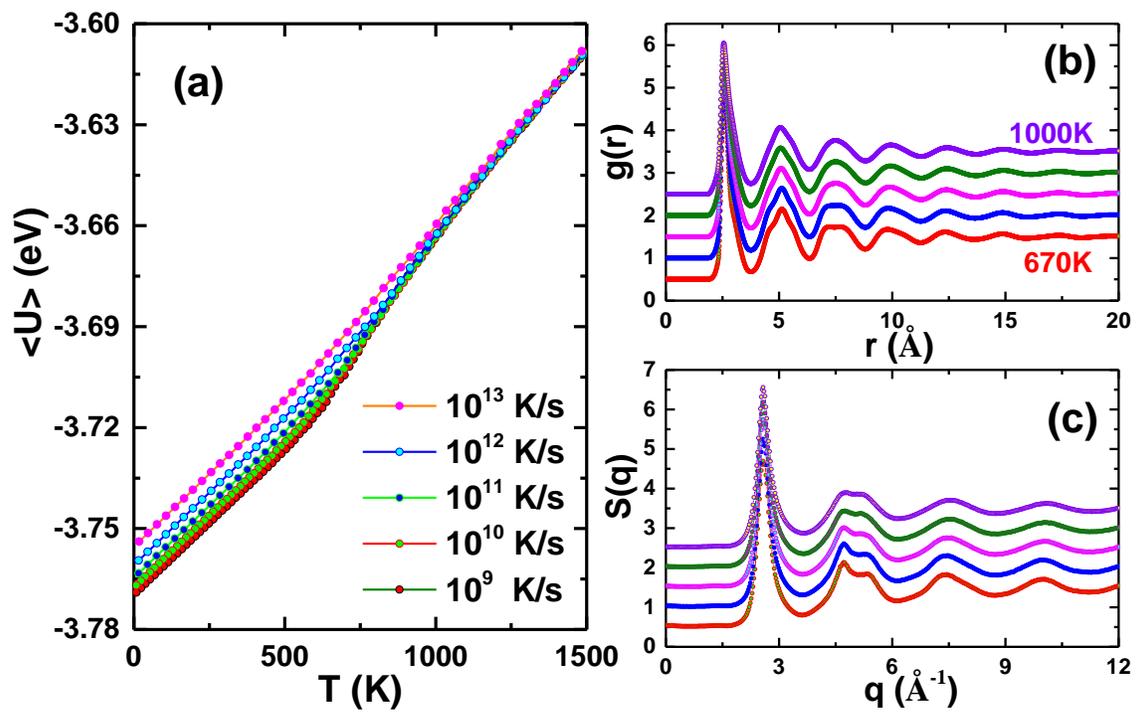

Figure 1. Hu et al.

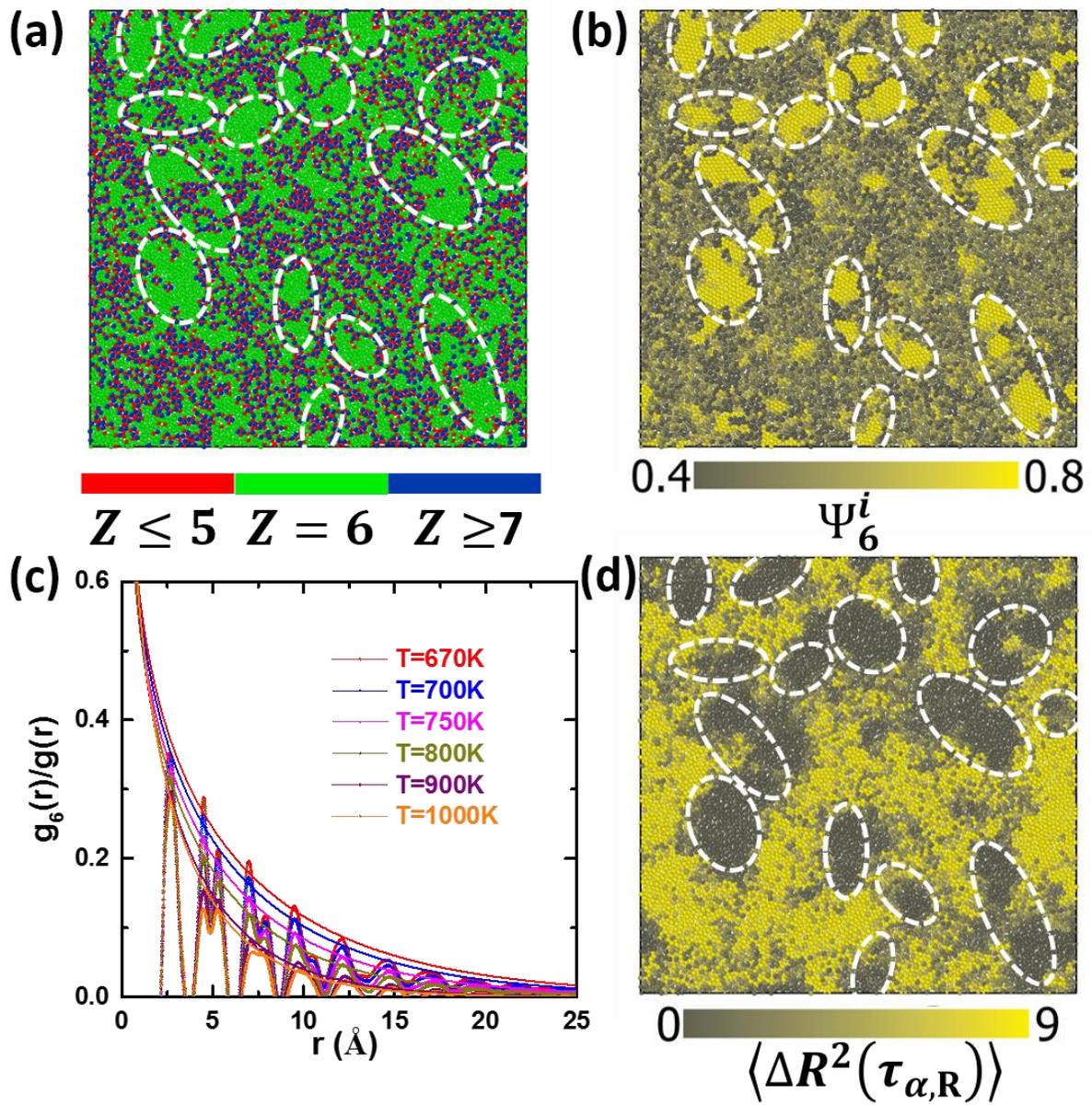

Figure 2. Hu et al.

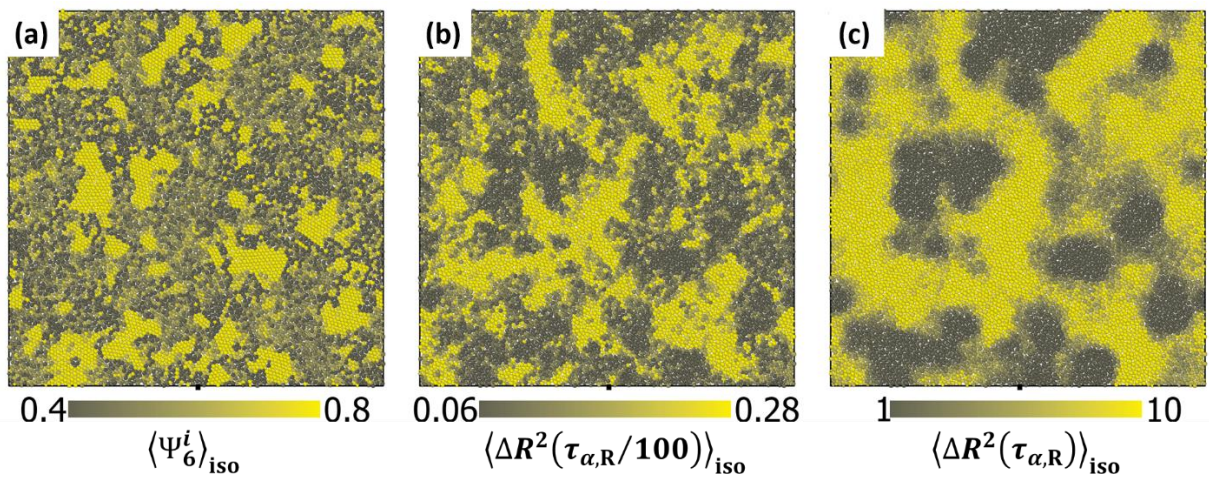

Figure 3. Hu et al.

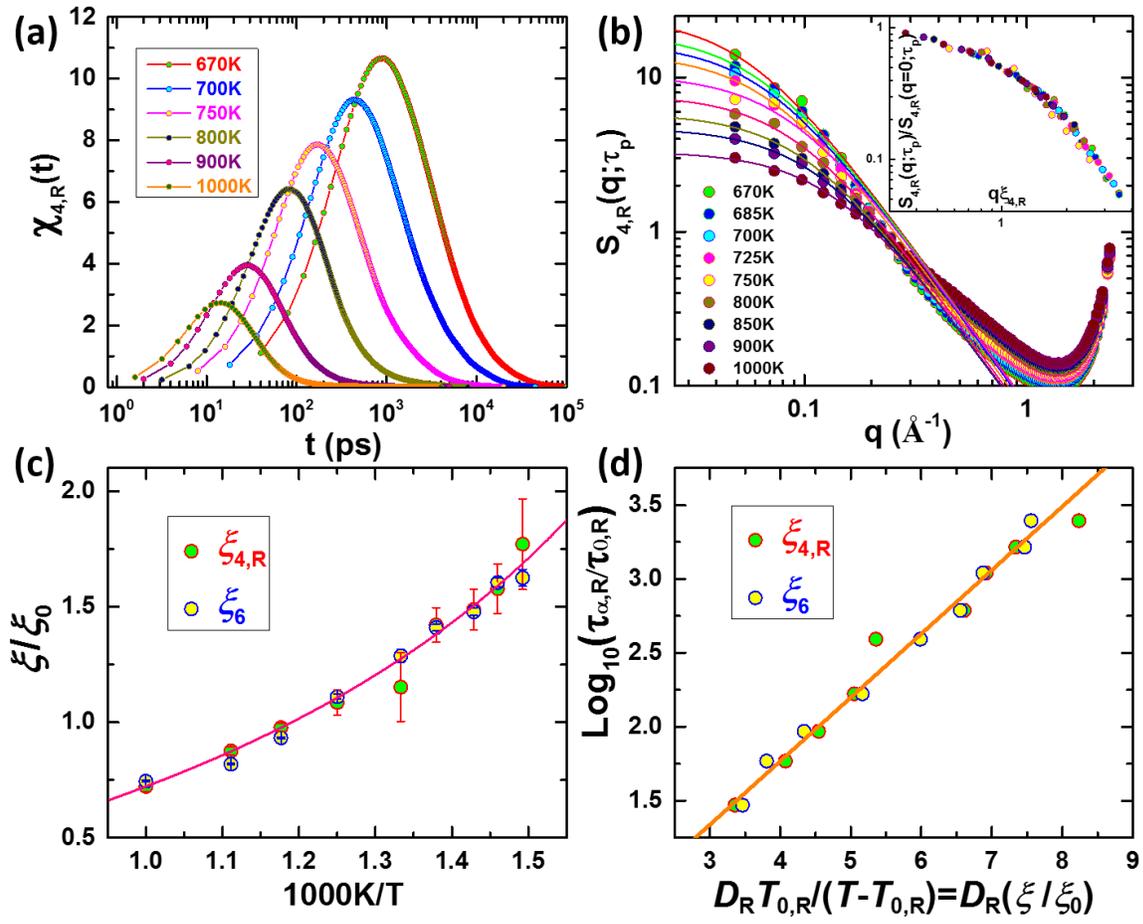

Figure 4. Hu et al.